\documentclass[aps,prl,reprint,superscriptaddress,floatfix]{revtex4-2}

\usepackage{amsmath,amssymb,bm}
\usepackage{graphicx}
\usepackage{xcolor}
\usepackage{booktabs}
\usepackage{multirow}
\usepackage{siunitx}
\usepackage[colorlinks=true,linkcolor=blue,citecolor=blue,urlcolor=blue,unicode=false]{hyperref}
\usepackage{microtype}
\usepackage{mathtools}
\usepackage[normalem]{ulem}

\newcommand{\MeV}{\,\mathrm{MeV}}
\begin{document}

\title{{Self-Consistent Subtracted Second Quasiparticle Random-Phase Approximation for Dipole Responses in Neutron-Rich Oxygen Isotopes}}

\author{M. J. Yang}
\affiliation{School of Science, Key Laboratory of High Performance Scientific Computation, Xihua University, Chengdu 610039, China}
\author{X. Lian}
\affiliation{College of Physics, Sichuan University, Chengdu 610065, China}
\author{C. L. Bai}
\email{bclphy@scu.edu.cn}
\affiliation{College of Physics, Sichuan University, Chengdu 610065, China}
\affiliation{Shanghai Research Center for Theoretical Nuclear Physics, NSFC and Fudan University, Shanghai 200438, China}
\author{C. L. Zhang}
\email{lei@bnu.edu.cn}
\affiliation{Beijing Normal University, Beijing, China}
\author{H. Sagawa}
\email{hiroyuki.sagawa@gmail.com}
\affiliation{Center for Mathematics and Physics, University of Aizu, Aizu-Wakamatsu, Fukushima 965-8560, Japan}
\affiliation{RIKEN Nishina Center, Wako 351-0198, Japan}
\author{H. Q. Zhang}
\affiliation{China Institute of Atomic Energy, Beijing 102413, China}

\begin{abstract} 
{Describing the nuclear dipole response in open-shell nuclei across wide energy scales, from low-lying states below the neutron separation energy to giant resonances
near 30\,MeV,  within a single microscopic framework is a long-standing open problem.
Here we establish, for the first time, a self-consistent subtracted second quasiparticle
random-phase approximation (SSQRPA) that simultaneously incorporates pairing
correlations (HF+BCS), explicit two-quasiparticle (2qp) and four-quasiparticle (4qp)
configurations in a large model space, the full forward- and backward-going QRPA
structure, and a subtraction procedure to eliminate double counting and ensure stability.
No existing framework unifies all of these ingredients in the quasiparticle basis.
We apply SSQRPA to the dipole excitations of neutron-rich oxygen isotopes
$^{18,20,22}$O, where the standard QRPA fails to simultaneously account for
high-energy giant dipole resonance (GDR) states near 22--25\,MeV, medium-energy
GDR states in the 10--18\,MeV region, and isolated low-lying $1^-$ states below
the neutron separation energy.
In SSQRPA, the dominant isovector (IV) GDR strength shifts downward by
approximately 4\,MeV into the experimentally observed medium-energy region,
while the compressional isoscalar (IS) GDR strength remains concentrated at higher
energies, in qualitative agreement with the observed electromagnetic dissociation
cross sections.
For the low-lying states, the root-mean-square deviation from experiment is
reduced from 5.31\,MeV (QRPA) to 0.35\,MeV (SSQRPA) in $^{18}$O, and from
4.07\,MeV to 0.74\,MeV in $^{20}$O, demonstrating that 4qp correlations are
essential for a simultaneous description of dipole excitations over a wide range
of excitation energies.
The SSQRPA provides a powerful and general framework for future microscopic
studies of collective excitations in open-shell nuclei across the nuclear mass table.}
\end{abstract}

\maketitle

{The nuclear dipole response ($J^\pi=1^-$) governs some of the most
fundamental processes in nuclear physics and astrophysics.
In the isovector (IV) channel, the giant dipole resonance (GDR) --- arising from
the out-of-phase collective oscillation of protons and neutrons --- dominates the
photonuclear cross section and strongly influences radiative neutron-capture rates
relevant to r-process nucleosynthesis~\cite{GORIELY2002217,BRACCO2019360,LANZA2023104006}.
Low-energy dipole states below the neutron separation energy, often associated
with pygmy resonances in neutron-rich systems, are sensitive to shell effects,
pairing, continuum coupling, and correlations beyond the two-quasiparticle level;
they may play a critical role in the r-process by enhancing neutron-capture cross
sections in neutron-rich nuclei~\cite{BRACCO2019360,SAVRAN2013210,LANZA2023104006}.
Complementarily, the compressional isoscalar (IS) dipole response provides
direct information on nuclear incompressibility and compressional
dynamics~\cite{GARG201855}.
A comprehensive theoretical description of the dipole response must therefore
account simultaneously for the GDR at high excitation energies, intermediate
fragmentation, and the fine structure of low-lying states --- a challenge that
spans nearly two decades in excitation energy. It requires a thoughtful treatment
of many-body correlations at multiple levels~\cite{RevModPhys.55.287,harakeh2001giant}.}

{The random-phase approximation (RPA) and its quasiparticle extension (QRPA)
provide the standard starting point for collective nuclear excitations, connecting
nuclear response functions to ground-state properties within density functional
theory~\cite{Bender2003,Onida2002,Chen2017}.
However, the restricted 2qp configuration space of QRPA captures only the leading
mean-field response and misses the spreading width, damping, and strength
redistribution generated by higher-order correlations~\cite{RevModPhys.55.287,harakeh2001giant}.
Various extensions have been developed to address these limitations.
In electronic systems, beyond-RPA correlations are incorporated through
configuration-space extensions such as second RPA~\cite{Peng2014} and through
frequency-dependent response kernels in many-body Green's-function
approaches~\cite{Sangalli2011,Bintrim2022}.
In nuclear physics, beyond-QRPA correlations can be incorporated via couplings to
collective phonons in quasiparticle-vibration coupling~\cite{PhysRevLett.131.082501,PhysRevC.109.044308},
multiphonon models~\cite{COLO2001427,PhysRevC.97.034311},
and quasiparticle time-blocking approaches~\cite{PhysRevC.78.014312,PhysRevC.75.054318},
as well as through correlated propagators in dressed RPA~\cite{PhysRevC.68.014311,PhysRevC.99.054327}.
The second RPA (SRPA) offers the most direct extension by explicitly enlarging the
configuration space to include the full set of $2p$-$2h$ excitations on top of the
Hartree-Fock ground state~\cite{PhysRevC.81.024317,PhysRevC.81.054312,GAMBACURTA2018163,PhysRevLett.125.212501,PhysRevC.106.014319}.
Self-consistent implementations of SRPA based on the Skyrme energy-density
functional (EDF), and in particular the subtracted SRPA (SSRPA) variant that removes
double counting of ground-state correlations and stabilizes the extended configuration
space~\cite{PhysRevC.88.054301,PhysRevC.92.034303}, have demonstrated that $2p$-$2h$
correlations are essential for a quantitative description of the spreading width and
fragmentation pattern of giant resonances (GR) in closed-shell nuclei~\cite{GAMBACURTA2026104251}.
Crucially, however, existing SRPA implementations are restricted to closed-shell nuclei,
where the ground state is well represented by a single Slater determinant.}

{For open-shell nuclei with significant pairing correlations, a natural and necessary
extension of SRPA must be formulated in the quasiparticle basis, so that the
higher-order sector consists of 4qp configurations.
Although 4qp correlations have been explored in earlier Tamm-Dancoff-type
approaches~\cite{Savoia1964FourQE,PhysRev.153.1138,PhysRev.155.1144,1968PhRv,PhysRevC.101.044314},
these retain the Tamm-Dancoff structure and therefore omit the backward-going
amplitudes and associated ground-state correlations of RPA.
To date, no self-consistent quasiparticle counterpart of SRPA has been formulated
that simultaneously incorporates pairing, explicit 2qp-4qp coupling, the full
forward- and backward-going QRPA structure, and a subtraction procedure to avoid
double counting and preserve stability~\cite{Engel_2017,GAMBACURTA2026104251}.
Here for the first time we establish such a framework, which we refer to as the
self-consistent subtracted second quasiparticle random-phase approximation
(SSQRPA).
This fills a fundamental gap in the hierarchy of nuclear many-body methods for
open-shell nuclei, providing a microscopic approach to correlated excitations from
low-lying states to giant resonances within a unified, self-consistent framework.}

{We demonstrate the SSQRPA by applying it to the dipole excitations of neutron-rich
oxygen isotopes $^{18,20,22}$O, which provide an ideal first testing ground because
their relatively small mass renders the 4qp configuration space computationally
tractable, while their dipole spectra exhibit the full complexity that challenges
existing models.
In these nuclei, dipole strength has been experimentally identified in three
distinct energy regions: (i) high-energy GDR states near 22--25\,MeV, (ii) medium-energy
GDR states from 10 to 18\,MeV whose electromagnetic dissociation cross sections
confirm their IV character~\cite{Aumann2001,PhysRevLett.86.5442}, and (iii) isolated
low-lying $1^-$ states below the neutron separation
energy~\cite{Tilley1995,TRYGGESTAD200252,NAKATSUKA2017387}.
Standard QRPA calculations yield IV giant-resonance strength predominantly near
20\,MeV~\cite{Bianco_2014}, far above the observed medium-energy region, and produce
no states below the neutron separation energy.
Even the QRPA-plus-phonon model, which incorporates selected higher-order
correlations, shifts some strength below 15\,MeV but leaves the dominant IV
contribution concentrated near 24\,MeV~\cite{COLO2001427}.
No existing approach achieves a simultaneous description of GDR fragmentation,
the medium-energy IV strength, and the low-lying $1^-$ state energies.
The SSQRPA results presented here resolve this long-standing problem, demonstrating
that the 4qp correlations treated within SSQRPA provide the essential mechanism for
a coherent description of dipole excitations over a wide excitation-energy range in
open-shell nuclei.}

{In the dipole excitations, the IV and compressional IS responses are the extensively studied modes,
which are excited by the operators}:
\begin{align}
\hat O^{\mathrm{IV}}_{1\mu}={}&\frac{eN}{A}\sum_{p=1}^{Z}r_pY_{1\mu}(\hat r_p)-\frac{eZ}{A}\sum_{n=1}^{N}r_nY_{1\mu}(\hat r_n),\label{eq:ivop}\\
\hat O^{\mathrm{IS}}_{1\mu}={}&\sum_{i=1}^{A}\left(r_i^3-\frac{5}{3}\langle r^2\rangle r_i\right)Y_{1\mu}(\hat r_i).\label{eq:isop}
\end{align}

The SQRPA creation operator explicitly contains both 2qp and 4qp components,
\begin{equation}
\begin{aligned}
Q_\nu^\dagger ={}& \sum_{\alpha\beta}\bigl(X_{\alpha\beta}^\nu\alpha_\alpha^\dagger\alpha_\beta^\dagger
-Y_{\alpha\beta}^\nu\alpha_\beta\alpha_\alpha\bigr)\\
&+\sum_{\alpha\beta\gamma\delta}\bigl(X_{\alpha\beta\gamma\delta}^\nu
\alpha_\alpha^\dagger\alpha_\beta^\dagger\alpha_\gamma^\dagger\alpha_\delta^\dagger\\
&\hspace{4em}-Y_{\alpha\beta\gamma\delta}^\nu\alpha_\delta\alpha_\gamma\alpha_\beta\alpha_\alpha\bigr),
\end{aligned}
\label{eq:operator}
\end{equation}
where Greek subscripts $\alpha,\beta,\gamma$, and $\delta$ denote single-quasiparticle states.  {The SQRPA equations are formulated as}
\begin{equation}
\begin{pmatrix}A&B\\-B^*&-A^*\end{pmatrix}
\begin{pmatrix}X^\nu\\Y^\nu\end{pmatrix}
=\hbar\omega_\nu
\begin{pmatrix}X^\nu\\Y^\nu\end{pmatrix},
\label{eq:SSQRPA}
\end{equation}
with
\begin{equation}
A=\begin{pmatrix}A_{22}&A_{24}\\A_{42}&A_{44}\end{pmatrix},\qquad
B=\begin{pmatrix}B_{22}&B_{24}\\B_{42}&B_{44}\end{pmatrix},
\label{eq:blocks}
\end{equation}
where indices 2 and 4 refer to the 2qp and 4qp sectors, respectively. The explicit matrix elements are given in the Supplemental Material~\cite{SupplementalMaterial}.

\begin{figure}[tbp]
\centering
\includegraphics[width=\linewidth]{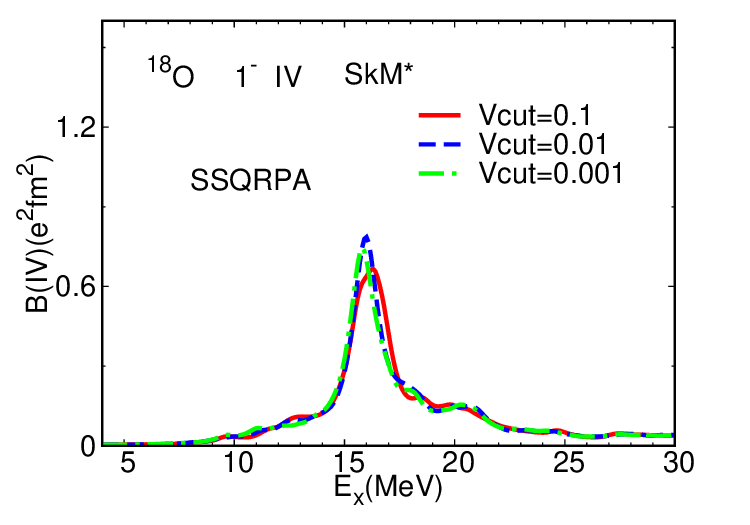}
\caption{IV dipole strength in $^{18}$O calculated with SSQRPA for three values of the 4qp cutoff $V_{\mathrm{cut}}$.
The discrete transition strengths are folded with a Lorentzian of width $\Gamma=1\MeV$. }
\label{fig:vcut}
\end{figure}

As the first step of the calculation, the  HF+BCS ground state is calculated in a harmonic-oscillator basis with $N_{\max}=12$ and $J_{\max}=11/2$. The SkM*  {EDF}~\cite{BARTEL198279} is employed in the particle-hole channel.
This parametrization gives a good overall description of giant resonance systematics, including the IV dipole response~\cite{GLEISSL1990205}, and its nuclear-matter incompressibility of $\rm K_\infty \simeq$ 217 MeV makes it a reasonable choice for studying the IS compressional dipole response~\cite{COLO2000362}. A volume pairing interaction is used in the particle-particle (pp) channel, with its strength adjusted to reproduce the empirical neutron pairing gap {of each oxygen isotope} obtained from the three-point formula~\cite{PhysRevLett.81.3599,CHANGIZI2015210}.
The 2qp configurations are built from the quasiparticle states with equivalent HF energy below 60 MeV, requiring either $|\rm U_{\alpha} V_{\beta} |$ or $|\rm V_{\alpha} U_{\beta} |$ larger than 10$^{-3}$,  where U and V denote the BCS vacancy and occupation amplitudes, respectively. The convergence of the 2qp configuration space with respect to these cutoffs has been explicitly verified.
To make the 4qp configuration space computationally tractable, we introduce an occupation cutoff $V_{\mathrm{cut}}$, in which quasiparticle states  with $V^2>1-V_{\mathrm{cut}}$ (
$V^2<V_{\mathrm{cut}}$)  are classified as hole-like (particle-like) states, respectively.
Under this classification, the configurations of 2p-2h type are included in the SQRPA  {configuration} space, but 4p, 4h, 3p1h, and 3h1p types are excluded.
Furthermore, to avoid double counting of correlations already encoded in the energy-density functional and to stabilize the extended configuration-space calculation, the subtraction prescription (SSQRPA) is applied~\cite{PhysRevC.88.054301,PhysRevC.92.034303}.  The necessity of the subtraction procedure is illustrated by a numerical example in the Supplemental Material.
This procedure was also found to stabilize solutions in the enlarged configuration space in Skyrme-SRPA calculations~\cite{PhysRevC.92.034303}.
In calculating the SSQRPA matrices, for the 2qp sector, both particle-hole (ph) and pp residual-interaction channels are included,
whereas only the ph channel is retained in the 4qp sector as the pp contributions are not explicitly separated in the present implementation {(see Supplemental Material for details)}.

\begin{table}[tbp]
\centering
\caption{Energy moments $m_1$ and $m_{-1}$  of the $^{18}$O IV dipole response calculated with QRPA and SSQRPA with different $V_{\mathrm{cut}}$. The classical TRK sum rule~\cite{Sandler1980TRK} is also shown for reference.}
\label{tab:moments}
\begin{tabular}{lccccc}
\toprule
 & TRK & QRPA & $V_{\mathrm{cut}}$=0.1 & $V_{\mathrm{cut}}$=0.01 &  $V_{\mathrm{cut}}$= 0.001\\
\midrule
$m_1$ & 66.01 & 84.67 & 120.8 & 121.3 & 123.8\\
$m_{-1}$ & -- & 0.197 & 0.197 & 0.197 & 0.197\\
\bottomrule
\end{tabular}
\end{table}

To assess the convergence with respect to the truncation of the 4qp space, the IV  dipole strength distributions of $^{18}$O are calculated
 with $V_{\mathrm{cut}}=0.1$, 0.01, and 0.001, and the results are shown in Fig.~\ref{fig:vcut}.
As shown in Fig.~\ref{fig:vcut}, the strength distributions obtained with $V_{\mathrm{cut}}=0.01$ and 0.001 show only minor deviations, and the energies of their main peaks differ by less than 0.1 MeV.
The energy moments $m_k=\sum_{\nu} E^k_{\nu} B_\nu$ provide an additional measure of the numerical convergence and characterize the global response {of SSQRPA}. As shown in Table~\ref{tab:moments}, the QRPA value of $m_1$ exceeds the classical Thomas-Reiche-Kuhn (TRK) sum rule, as expected for Skyrme energy-density functionals because the momentum-dependent terms generate the TRK enhancement~\cite{Bohigas1979,Oishi2016}. The subtraction procedure further increases $m_1$ in SSQRPA,
whereas the inverse-energy-weighted moment $m_{-1}$ remains essentially unchanged from QRPA to SSQRPA. This behavior is consistent with the {conservation}  of the static response under the subtraction procedure, as previously found in SRPA~\cite{PhysRevC.92.034303}.
Moreover, reducing $V_{\mathrm{cut}}$ from $0.01$ to $0.001$ approximately doubles the 4qp configuration space,
while the corresponding energy-weighted moment $m_1$ changes by only about 2\% and $m_{-1}$ remains unchanged.
We therefore adopt $V_{\mathrm{cut}}=0.01$ for the following calculations.

\begin{figure}[tbp]
\centering
\includegraphics[width=\linewidth]{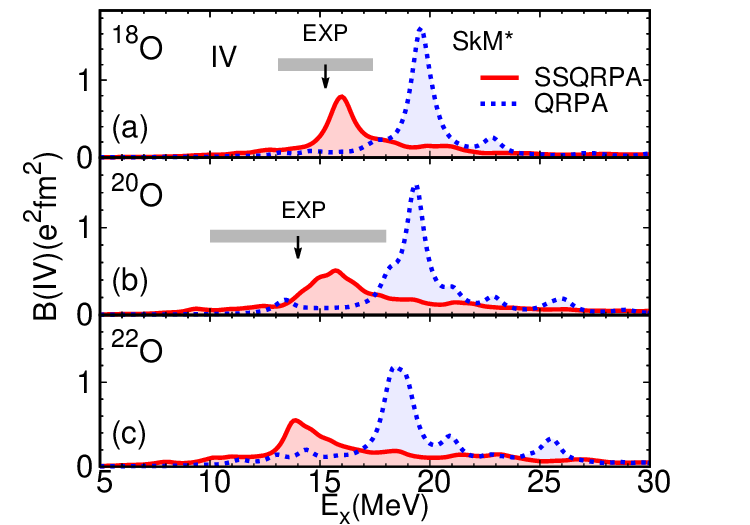}
\caption{IV dipole strength distributions of $^{18,20,22}$O in the excitation-energy range $5$--$30\MeV$. The blue dotted and red solid curves denote the QRPA and SSQRPA results, respectively. In $^{18}$O [panel (a)], the shaded interval from $13.1$ to $17.4\MeV$ indicates the range of experimentally identified $1^-$ resonances~\cite{Tilley1995}. The shaded region in  $^{20}$O [panel (b)] indicates the excitation-energy
interval in which a pronounced electromagnetic excitation cross section was
observed in Coulomb-excitation measurements
\cite{Aumann2001}. The arrow represents the midpoint of the energy interval.}
\label{fig:iv}
\end{figure}

\begin{figure}[tbp]
\centering
\includegraphics[width=\linewidth]{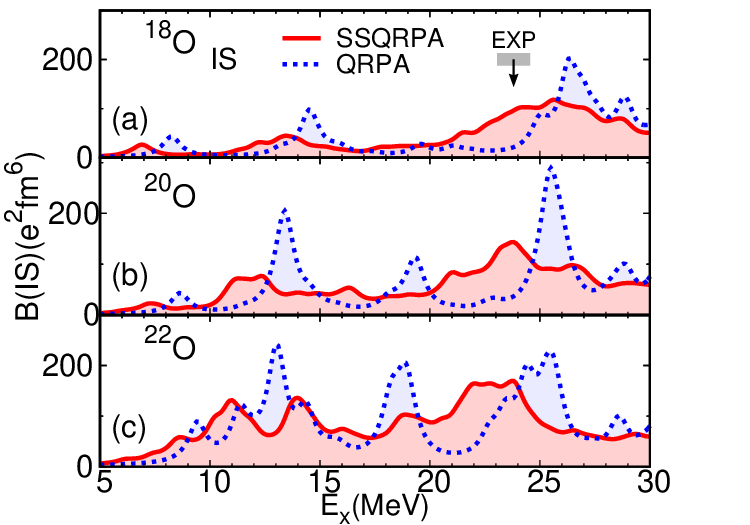}
\caption{IS compressional dipole strength distributions of $^{18,20,22}$O in the excitation-energy range $5$--$30\MeV$. The blue dotted and red solid curves show the QRPA and SSQRPA results, respectively. The shaded interval from $23$ to $25\MeV$ in  $^{18}$O indicates the range of experimentally identified $1^-$ resonances~\cite{Tilley1995}. }
\label{fig:is}
\end{figure}

The IV  dipole strength distributions calculated by QRPA and SSQRPA are shown in Fig.~\ref{fig:iv}.
The principal effect of SSQRPA on the IV dipole response is a substantial downward shift of the dominant IV dipole strength toward the experimentally observed energy region.
In the QRPA calculation, the IV dipole strength is predominantly concentrated in a peak at $E_x\sim19.5\MeV$ in $^{18,20}$O and at $E_x\sim18.5\MeV$ in $^{22}$O.
For $^{18}$O, the main peak in the SSQRPA calculation is shifted downward by about 4 MeV into the experimental GDR energy region extending from $13.1$ to $17.4\MeV$~\cite{Tilley1995}.
For $^{20}$O,  Coulomb-excitation measurements reveal sizable dipole strength in the energy region 10--18 MeV~\cite{Aumann2001}.
This feature is absent in the QRPA result, whereas in SSQRPA the dominant IV dipole strength is shifted downward to the 15--16 MeV energy region.
As a result, SSQRPA predicts IV GDR strength concentrated in the medium energy region, in qualitative and quantitative agreement with experimental measurements.

It is known that the high-energy branch of the ISGDR has been associated with compressional motion~\cite{Repko2013}.
 We calculate the IS compressional dipole response within SSQRPA to explore possible isoscalar properties in these states, and the results are shown in Fig.~\ref{fig:is}.
 In $^{18}$O, the strength obtained by QRPA is mainly concentrated in the 25--27 MeV region. In the SSQRPA calculation, however, the strength becomes strongly fragmented and is redistributed toward the 23--27 MeV region, which overlaps with the experimental $1^-$ energy region~\cite{Tilley1995}.
 A similar trend is found in $^{20}$O, where the centroids of the broad high-energy structures shift from 25.5 MeV in QRPA to 23.8 MeV in SSQRPA (from 24.4 MeV to 23.8 MeV for $^{22}$O), while the strength distributions become substantially more fragmented.
The SSQRPA calculations suggest possible IS compressional dipole components in the high-energy GDR region. Appreciable IS strength is also predicted at intermediate and low excitation energies, including the region below the neutron separation energy.

\begin{figure}[tbp]
\centering
\includegraphics[width=0.95\linewidth]{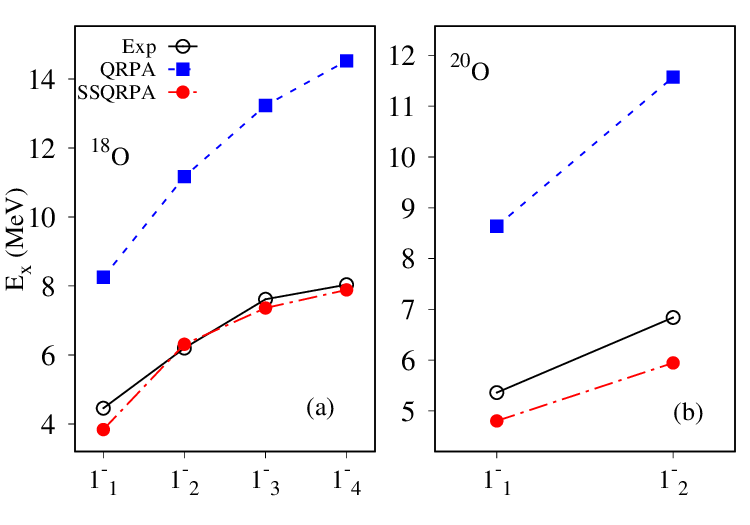}
\caption{Low-energy $1^-$ excitation energies in $^{18}$O (left) and $^{20}$O (right). Experimental levels are shown by open circles, while QRPA and SSQRPA results are shown by blue squares and red circles, respectively.  The experimental data are taken from Refs~\cite{Tilley1995,TRYGGESTAD200252,NAKATSUKA2017387}.}
\label{fig:lowlying}
\end{figure}

Besides the giant resonances, low-lying states below the neutron separation energy provide a sensitive probe of the interplay between
single-particle structure and many-body correlations,
making their excitation energies a stringent benchmark for theoretical approaches that include correlations beyond the 2qp space.
The excitation energies of the low-lying $1^-$ states in $^{18}$O and $^{20}$O are shown in Fig.~\ref{fig:lowlying}.
As shown in the figure, QRPA systematically overestimates the excitation energies by about 4-5 MeV and does not produce states below the neutron separation energy, whereas SSQRPA reproduces the experimental excitation energies well, with deviations smaller than 1 MeV. The root-mean-square (rms) deviation is reduced from 5.31 MeV in QRPA to 0.35 MeV in SSQRPA for the four lowest states in $^{18}$O,  and from 4.07 MeV in QRPA to 0.74 MeV in SSQRPA for the two measured states in $^{20}$O.
{We note that the absolute dipole strengths of the low-lying states remain one to two orders of magnitude smaller than the experimental values, suggesting that additional higher-order correlations, such as continuum coupling and beyond-4qp configurations, may contribute to the transition strengths and warrant future investigation.}

In summary, we have developed for the first time a state-of-the-art SSQRPA model that explicitly and completely includes 2qp and 4qp many-body correlations in a large model space and applied it to the dipole response of $^{18,20,22}$O.
In the calculations, SSQRPA simultaneously reproduces the excitation energies of low-energy dipole states, medium-energy GDR states, and the qualitative features of high-energy GDR states.
The calculations also suggest that the high-energy GDR states carry significant IS compressional character, and that the medium-energy GDR states exhibit a mixing of IV and IS compressional components.
These results demonstrate that the 4qp correlations incorporated within SSQRPA provide an essential mechanism for describing dipole excitations over widely separated energy scales in open-shell nuclei.
The self-consistent SSQRPA model now provides a powerful framework for future microscopic studies of open-shell nuclei across a wide area of the nuclear mass table, including various GR and pygmy excitations as well as charge-exchange Gamow-Teller states.  Double giant resonances will also be investigated in future work.

\begin{acknowledgments}
We thank Li Wanhui for fruitful discussions. This work was supported by the
National Natural Science Foundation of China under Grant No. 12575126 and 12547102, and by Xihua University under Grant No. ZX20260046.
{H.S. is supported by the JSPS Grant-in-Aid for Scientific Research (C) under Grant No. JP26K07079.}
\end{acknowledgments}

\nocite{Yannouleas1983MicroscopicDissipationII}
\bibliography{E1}

\end{document}


\title{Supplemental Material for ``Self-Consistent Subtracted Second Quasiparticle Random-Phase Approximation for Dipole Responses in Neutron-Rich Oxygen Isotopes”}
\author{M. J. Yang}
\affiliation{School of Science, Key Laboratory of High Performance Scientific Computation, Xihua University, Chengdu 610039, China}
\author{X. Lian}
\affiliation{College of Physics, Sichuan University, Chengdu 610065, China}
\author{C. L. Bai}
\email{bclphy@scu.edu.cn}
\affiliation{College of Physics, Sichuan University, Chengdu 610065, China}
\author{C. L. Zhang}
\email{lei@bnu.edu.cn}
\affiliation{Beijing Normal University, Beijing, China}
\author{H. Sagawa}
\email{hiroyuki.sagawa@gmail.com}
\affiliation{Center for Mathematics and Physics, University of Aizu, Aizu-Wakamatsu, Fukushima 965-8560, Japan}
\affiliation{RIKEN Nishina Center, Wako 351-0198, Japan}
\author{H. Q. Zhang}
\affiliation{China Institute of Atomic Energy, Beijing 102413, China}
\maketitle

\section{{I. Matrix elements of SQRPA}} 
The matrix elements for SQRPA are summarized in the explicit form. They are presented here rather than in the main text for brevity. The SQRPA matrix elements are obtained from the equation-of-motion and double commutator formulation in the BCS quasiparticle vacuum, which is analogous to the standard SRPA formalism~\cite{Savoia1964FourQE,PhysRev.153.1138,PhysRev.155.1144,1968PhRv,PhysRevC.81.054312,Yannouleas1983MicroscopicDissipationII}.

We start from the original Hamiltonian with an antisymmetrized two-body interaction,
\begin{equation}\label{eqH}
 H=\sum_{ij}(t_{ij}-\lambda\delta_{ij})a_i^\dagger a_j
 +\frac{1}{4}\sum_{ijkl}V_{ij,kl}a_i^\dagger a_j^\dagger a_l a_k.
\end{equation}
Here $V_{ij,kl}$ is the antisymmetrized two-body matrix element, and all particle-state indices are summed without restrictions.
Applying the Bogoliubov transformation,
\begin{equation}\label{eqBogoliubov}
\begin{aligned}
 a_\alpha^\dagger&=U_\alpha\alpha_\alpha^\dagger+V_\alpha\alpha_{\bar\alpha},&
 a_{\bar\alpha}^\dagger&=U_\alpha\alpha_{\bar\alpha}^\dagger-V_\alpha\alpha_\alpha,\\
 a_\alpha&=U_\alpha\alpha_\alpha+V_\alpha\alpha_{\bar\alpha}^\dagger,&
 a_{\bar\alpha}&=U_\alpha\alpha_{\bar\alpha}-V_\alpha\alpha_\alpha^\dagger,
\end{aligned}
\end{equation}
and normal ordering with respect to the BCS quasiparticle vacuum gives components $H^{mn}$, where $m$ and $n$ count quasiparticle creation and annihilation operators. The BCS stationarity condition sets $H^{20}=H^{02}=0$. After dropping the vacuum-energy constant, which does not enter the commutators, we write the transformed Hamiltonian as
\begin{equation}\label{eqHqp}
 H=H^{11}+H^{22}+H^{13}+H^{31}+H^{40}+H^{04}.
\end{equation}

For two- and four-quasiparticle configurations, define
\begin{equation}\label{eqConfigOps}
\begin{aligned}
 P_{\alpha\beta}^\dagger&=\alpha_\alpha^\dagger\alpha_\beta^\dagger,&
 P_{\alpha\beta}&=\alpha_\beta\alpha_\alpha,\\
 R_{\alpha\beta\gamma\delta}^\dagger
 &=\alpha_\alpha^\dagger\alpha_\beta^\dagger\alpha_\gamma^\dagger\alpha_\delta^\dagger,&
 R_{\alpha\beta\gamma\delta}
 &=\alpha_\delta\alpha_\gamma\alpha_\beta\alpha_\alpha.
\end{aligned}
\end{equation}
The SQRPA excitation operator is denoted by $Q_\nu^\dagger$ and contains forward and backward amplitudes in both sectors:
\begin{equation}\label{eqQphonon}
\begin{aligned}
 Q_\nu^\dagger={}&\sum_{\alpha\beta}
   (X_{\alpha\beta}^\nu P_{\alpha\beta}^\dagger-Y_{\alpha\beta}^\nu P_{\alpha\beta})\\
 &+\sum_{\alpha\beta\gamma\delta}
   (X_{\alpha\beta\gamma\delta}^\nu R_{\alpha\beta\gamma\delta}^\dagger
    -Y_{\alpha\beta\gamma\delta}^\nu R_{\alpha\beta\gamma\delta}).
\end{aligned}
\end{equation}
The commutator entering the equation of motion is
\begin{equation}\label{eqHQcomm}
\begin{aligned}
 [H,Q_\nu^\dagger]={}&\sum_{\alpha\beta}
   (X_{\alpha\beta}^\nu[H,P_{\alpha\beta}^\dagger]
    -Y_{\alpha\beta}^\nu[H,P_{\alpha\beta}])\\
 &+\sum_{\alpha\beta\gamma\delta}
   (X_{\alpha\beta\gamma\delta}^\nu[H,R_{\alpha\beta\gamma\delta}^\dagger]
    -Y_{\alpha\beta\gamma\delta}^\nu[H,R_{\alpha\beta\gamma\delta}]).
\end{aligned}
\end{equation}
Projecting this equation onto the two- and four-quasiparticle configurations gives the following contributions of the transformed original Hamiltonian to the SQRPA blocks:
\begin{equation}\label{eqABsectors}
\begin{aligned}
 A_{IJ}&=\langle0|[P_I,[H^{11}+H^{22},P_J^\dagger]]|0\rangle,
 && I,J\in 2qp,\\
 A_{IK}&=\langle0|[P_I,[H^{13},R_K^\dagger]]|0\rangle,
 && I\in2qp,\ K\in4qp,\\
 A_{KL}&=\langle0|[R_K,[H^{11}+H^{22},R_L^\dagger]]|0\rangle,
 && K,L\in4qp,\\
 B_{IJ}&=-\langle0|[P_I,[H^{40},P_J]]|0\rangle,
 && I,J\in2qp,\\
 B_{IK}&=-\langle0|[P_I,[H,R_K]]|0\rangle=0,
 && I\in2qp,\ K\in4qp,\\
 B_{KL}&=-\langle0|[R_K,[H,R_L]]|0\rangle=0,
 && K,L\in4qp.
\end{aligned}
\end{equation}
Here $I,J$ label two-quasiparticle configurations and $K,L$ label four-quasiparticle configurations. The mixed and four-quasiparticle $B$ blocks vanish because the original two-body Hamiltonian generates at most four quasiparticles, while the corresponding vacuum matrix elements require six and eight, respectively. The explicit matrix elements are given next.

The matrix elements in  the SQRPA matrices in Eqs. (4) and (5) in the main text are written as
\begin{equation}\label{eqA22}
	\begin{split}		
		A_{22}=&A_{\alpha\beta,\alpha^\prime\beta^\prime}=<|[\alpha_\beta\alpha_\alpha,H,\alpha_{\alpha^\prime}^\dagger\alpha_{\beta^\prime}^\dagger]|>\\
		=&(E_{\alpha}+E_{\beta})\delta_{\alpha\alpha^\prime}\delta_{\beta\beta^\prime}\\
		+&V^{pp}_{\alpha\beta\alpha^\prime\beta^\prime}(U_\alpha U_\beta U_{\alpha^\prime}U_{\beta^\prime}+ V_{\alpha}V_\beta V_{\alpha^\prime}V_{\beta^\prime})\\
		+&V_{\alpha\bar{\beta^\prime}\bar{\beta}\alpha^\prime}(U_\alpha V_\beta U_{\alpha^\prime}V_{\beta^\prime}+V_\alpha U_\beta V_{\alpha^\prime}U_{\beta^\prime})\\
		-&V_{\alpha\bar{\alpha^\prime}\bar{\beta}\beta^\prime}(U_\alpha V_\beta V_{\alpha^\prime}U_{\beta^\prime}+V_\alpha U_\beta U_{\alpha^\prime}V_{\beta^\prime})
	\end{split}
\end{equation}

\begin{equation}\label{eqB22}
	\begin{split}		
		B_{22}=&B_{\alpha\beta,\alpha^\prime\beta^\prime}=-<|[\alpha_\beta\alpha_\alpha,H,\alpha_{\beta^\prime}\alpha_{\alpha^\prime}]|>\\
		=&-V^{pp}_{\alpha\beta\bar{\alpha^\prime}\bar{\beta^\prime}}(U_\alpha U_\beta V_{\alpha^\prime}V_{\beta^\prime} + V_{\alpha}V_\beta U_{\alpha^\prime}U_{\beta^\prime})\\
		&+V_{\alpha\alpha^\prime\bar{\beta}\bar{\beta^\prime}} (U_{\alpha}V_\beta U_{\alpha^\prime}V_{\beta^\prime} +V_\alpha U_\beta V_{\alpha^\prime}U_{\beta^\prime})\\
		& - V_{\alpha\beta^\prime\bar{\beta}\bar{\alpha^\prime}} (U_\alpha V_\beta V_{\alpha^\prime}U_{\beta^\prime} +V_\alpha U_\beta U_{\alpha^\prime} V_{\beta^\prime})
	\end{split}
\end{equation}

\begin{equation}\label{eqA24}
	\begin{split}
		&A_{24}=A_{\alpha\beta,\alpha^\prime\beta^\prime\gamma^\prime\delta^\prime}=<|[\alpha_\beta\alpha_\alpha,H,\alpha_{\alpha^\prime}^\dagger\alpha_{\beta^\prime}^\dagger\alpha_{\gamma^\prime}^\dagger\alpha_{\delta^\prime}^\dagger]|>\\
		&=(1-P_{\alpha\beta})\{\quad \\
		& -\delta_{\beta\delta^\prime}\textbf{(}  V_{\alpha\bar{\gamma^\prime}\alpha^\prime\beta^\prime}(U_\alpha V_{\gamma^\prime}U_{\alpha^\prime}U_{\beta^\prime}-V_\alpha U_{\gamma^\prime}V_{\alpha^\prime}V_{\beta^\prime})\\ 
		&\qquad +V_{\alpha\bar{\alpha^\prime}\beta^\prime\gamma^\prime}(U_{\alpha}V_{\alpha^\prime}U_{\beta^\prime}U_{\gamma^\prime}-V_{\alpha}U_{\alpha^\prime}V_{\beta^\prime}V_{\gamma^\prime}) \\
		&\qquad +V_{\alpha\bar{\beta^\prime}\gamma^\prime\alpha^\prime}(U_\alpha V_{\beta^\prime}U_{\gamma^\prime}U_{\alpha^\prime}-V_\alpha U_{\beta^\prime}V_{\gamma^\prime}V_{\alpha^\prime}) \textbf{)} \\ 
		&+\delta_{\beta\gamma^\prime}\textbf{(} V_{\alpha\bar{\delta^\prime}\alpha^\prime\beta^\prime}(U_\alpha V_{\delta^\prime}U_{\alpha^\prime}U_{\beta^\prime}-V_\alpha U_{\delta^\prime}V_{\alpha^\prime}V_{\beta^\prime})\\
		&\qquad +V_{\alpha\bar{\alpha^\prime}\beta^\prime\delta^\prime}(U_{\alpha}V_{\alpha^\prime}U_{\beta^\prime}U_{\delta^\prime}-V_{\alpha}U_{\alpha^\prime}V_{\beta^\prime}V_{\delta^\prime})\\
		&\qquad + V_{\alpha\bar{\beta^\prime}\delta^\prime\alpha^\prime}(U_\alpha V_{\beta^\prime}U_{\delta^\prime}U_{\alpha^\prime}-V_\alpha U_{\beta^\prime}V_{\delta^\prime}V_{\alpha^\prime}) \textbf{)}\\
		&-\delta_{\beta\beta^\prime}\textbf{(} V_{\alpha\bar{\delta^\prime}\alpha^\prime\gamma^\prime}(U_\alpha V_{\delta^\prime}U_{\alpha^\prime}U_{\gamma^\prime}-V_\alpha U_{\delta^\prime}V_{\alpha^\prime}V_{\gamma^\prime})\\
		&\qquad + V_{\alpha\bar{\alpha^\prime}\gamma^\prime\delta^\prime}(U_{\alpha}V_{\alpha^\prime}U_{\gamma^\prime}U_{\delta^\prime}-V_{\alpha}U_{\alpha^\prime}V_{\gamma^\prime}V_{\delta^\prime})\\
		&\qquad+V_{\alpha\bar{\gamma^\prime}\delta^\prime\alpha^\prime}(U_\alpha V_{\gamma^\prime}U_{\delta^\prime}U_{\alpha^\prime}-V_\alpha U_{\gamma^\prime}V_{\delta^\prime}V_{\alpha^\prime}) \textbf{)}\\
		&+\delta_{\beta\alpha^\prime}\textbf{(} V_{\alpha\bar{\delta^\prime}\beta^\prime\gamma^\prime}(U_\alpha V_{\delta^\prime}U_{\beta^\prime}U_{\gamma^\prime}-V_\alpha U_{\delta^\prime}V_{\beta^\prime}V_{\gamma^\prime})\\
		&\qquad +V_{\alpha\bar{\beta^\prime}\gamma^\prime\delta^\prime}(U_{\alpha}V_{\beta^\prime}U_{\gamma^\prime}U_{\delta^\prime}-V_{\alpha}U_{\beta^\prime}V_{\gamma^\prime}V_{\delta^\prime})\\
		&\qquad +V_{\alpha\bar{\gamma^\prime}\delta^\prime\beta^\prime}(U_\alpha V_{\gamma^\prime}U_{\delta^\prime}U_{\beta^\prime}-V_\alpha U_{\gamma^\prime}V_{\delta^\prime}V_{\beta^\prime}) \textbf{)}\\
		&\qquad \quad\} 
	\end{split}
\end{equation}

\begin{equation}\label{eqA44t}
	\begin{split}	
		A_{44}=&A_{\alpha\beta\gamma\delta,\alpha^\prime\beta^\prime\gamma^\prime\delta^\prime}=<|[\alpha_\delta\alpha_\gamma\alpha_\beta\alpha_\alpha,H,\alpha_{\alpha^\prime}^\dagger\alpha_{\beta^\prime}^\dagger\alpha_{\gamma^\prime}^\dagger\alpha_{\delta^\prime}^\dagger]|>\\
		=&A^E_{\alpha\beta\gamma\delta,\alpha^\prime\beta^\prime\gamma^\prime\delta^\prime}+A^{V1}_{\alpha\beta\gamma\delta,\alpha^\prime\beta^\prime\gamma^\prime\delta^\prime}+A^{V2}_{\alpha\beta\gamma\delta,\alpha^\prime\beta^\prime\gamma^\prime\delta^\prime}
	\end{split}
\end{equation}
And 
\begin{equation}\label{A44e}
	A^E_{\alpha\beta\gamma\delta,\alpha^\prime\beta^\prime\gamma^\prime\delta^\prime}=(E_{\alpha}+E_{\beta}+E_{\gamma}+E_{\delta})\delta_{\alpha\alpha^\prime}\delta_{\beta\beta^\prime}\delta_{\gamma\gamma^\prime}\delta_{\delta\delta^\prime}
\end{equation}

\begin{align}\label{eqA22v1}
&A^{V1}_{\alpha\beta\gamma\delta,\alpha^\prime\beta^\prime\gamma^\prime\delta^\prime}\notag\\
=&(1-P_{\delta\gamma})\{ \notag\\ &V_{\alpha\beta\alpha^\prime\beta^\prime}\delta_{\gamma\gamma^\prime}\delta_{\delta\delta^\prime}(U_{\alpha}U_{\beta}U_{\alpha^\prime}U_{\beta^\prime}+V_{\alpha}V_{\beta}V_{\alpha^\prime}V_{\beta^\prime})\notag\\
		+&V_{\alpha\beta\gamma^\prime\delta^\prime}\delta_{\gamma\alpha^\prime}\delta_{\delta\beta^\prime}(U_{\alpha}U_{\beta}U_{\gamma^\prime}U_{\delta^\prime}+V_{\alpha}V_{\beta}V_{\gamma^\prime}V_{\delta^\prime})\notag\\
		-&V_{\alpha\beta\alpha^\prime\gamma^\prime}\delta_{\gamma\beta^\prime}\delta_{\delta\delta^\prime}(U_{\alpha}U_{\beta}U_{\alpha^\prime}U_{\gamma^\prime}+V_{\alpha}V_{\beta}V_{\alpha^\prime}V_{\gamma^\prime})\notag\\
		-&V_{\alpha\beta\beta^\prime\delta^\prime}\delta_{\gamma\alpha^\prime}\delta_{\delta\gamma^\prime}(U_{\alpha}U_{\beta}U_{\beta^\prime}U_{\delta^\prime}+V_{\alpha}V_{\beta}V_{\beta^\prime}V_{\delta^\prime})\notag\\
		+&V_{\alpha\beta\alpha^\prime\delta^\prime}\delta_{\gamma\beta^\prime}\delta_{\delta\gamma^\prime}(U_{\alpha}U_{\beta}U_{\alpha^\prime}U_{\delta^\prime}+V_{\alpha}V_{\beta}V_{\alpha^\prime}V_{\delta^\prime})\notag\\
		+&V_{\alpha\beta\beta^\prime\gamma^\prime}\delta_{\gamma\alpha^\prime}\delta_{\delta\delta^\prime}(U_{\alpha}U_{\beta}U_{\beta^\prime}U_{\gamma^\prime}+V_{\alpha}V_{\beta}V_{\beta^\prime}V_{\gamma^\prime}) \quad \}\notag\\ 
		+&(1-P_{\alpha\beta})\{ \notag\\
		& V_{\gamma\delta\alpha^\prime\beta^\prime}\delta_{\alpha\gamma^\prime}\delta_{\beta\delta^\prime}(U_{\gamma}U_{\delta}U_{\alpha^\prime}U_{\beta^\prime}+V_{\gamma}V_{\delta}V_{\alpha^\prime}V_{\beta^\prime})\notag\\
		+&V_{\gamma\delta\gamma^\prime\delta^\prime}\delta_{\alpha\alpha^\prime}\delta_{\beta\beta^\prime}(U_{\gamma}U_{\delta}U_{\gamma^\prime}U_{\delta^\prime}+V_{\gamma}V_{\delta}V_{\gamma^\prime}V_{\delta^\prime})\notag\\
		-&V_{\gamma\delta\alpha^\prime\gamma^\prime}\delta_{\alpha\beta^\prime}\delta_{\beta\delta^\prime}(U_{\gamma}U_{\delta}U_{\alpha^\prime}U_{\gamma^\prime}+V_{\gamma}V_{\delta}V_{\alpha^\prime}V_{\gamma^\prime})\notag\\
		-&V_{\gamma\delta\beta^\prime\delta^\prime}\delta_{\alpha\alpha^\prime}\delta_{\beta\gamma^\prime}(U_{\gamma}U_{\delta}U_{\beta^\prime}U_{\delta^\prime}+V_{\gamma}V_{\delta}V_{\beta^\prime}V_{\delta^\prime})\notag\\
		+&V_{\gamma\delta\alpha^\prime\delta^\prime}\delta_{\alpha\beta^\prime}\delta_{\beta\gamma^\prime}(U_{\gamma}U_{\delta}U_{\alpha^\prime}U_{\delta^\prime}+V_{\gamma}V_{\delta}V_{\alpha^\prime}V_{\delta^\prime})\notag\\
		+&V_{\gamma\delta\beta^\prime\gamma^\prime}\delta_{\alpha\alpha^\prime}\delta_{\beta\delta^\prime}(U_{\gamma}U_{\delta}U_{\beta^\prime}U_{\gamma^\prime}+V_{\gamma}V_{\delta}V_{\beta^\prime}V_{\gamma^\prime}) \quad\}\notag\\
		+&(1-P_{\alpha\gamma})\{ \notag\\
		-&V_{\beta\delta\alpha^\prime\beta^\prime}\delta_{\alpha\gamma^\prime}\delta_{\gamma\delta^\prime}(U_{\beta}U_{\delta}U_{\alpha^\prime}U_{\beta^\prime}+V_{\beta}V_{\delta}V_{\alpha^\prime}V_{\beta^\prime})\notag\\
		-&V_{\beta\delta\gamma^\prime\delta^\prime}\delta_{\alpha\alpha^\prime}\delta_{\gamma\beta^\prime}(U_{\beta}U_{\delta}U_{\gamma^\prime}U_{\delta^\prime}+V_{\beta}V_{\delta}V_{\gamma^\prime}V_{\delta^\prime})\notag\\
		+&V_{\beta\delta\alpha^\prime\gamma^\prime}\delta_{\alpha\beta^\prime}\delta_{\gamma\delta^\prime}(U_{\beta}U_{\delta}U_{\alpha^\prime}U_{\gamma^\prime}+V_{\beta}V_{\delta}V_{\alpha^\prime}V_{\gamma^\prime})\notag\\
		+&V_{\beta\delta\beta^\prime\delta^\prime}\delta_{\alpha\alpha^\prime}\delta_{\gamma\gamma^\prime}(U_{\beta}U_{\delta}U_{\beta^\prime}U_{\delta^\prime}+V_{\beta}V_{\delta}V_{\beta^\prime}V_{\delta^\prime})\notag\\
		-&V_{\beta\delta\alpha^\prime\delta^\prime}\delta_{\alpha\beta^\prime}\delta_{\gamma\gamma^\prime}(U_{\beta}U_{\delta}U_{\alpha^\prime}U_{\delta^\prime}+V_{\beta}V_{\delta}V_{\alpha^\prime}V_{\delta^\prime})\notag\\
		-&V_{\beta\delta\beta^\prime\gamma^\prime}\delta_{\alpha\alpha^\prime}\delta_{\gamma\delta^\prime}(U_{\beta}U_{\delta}U_{\beta^\prime}U_{\gamma^\prime}+V_{\beta}V_{\delta}V_{\beta^\prime}V_{\gamma^\prime}) \quad\}\notag\\
		+&(1-P_{\delta\beta})\{ \notag\\
		-&V_{\alpha\gamma\alpha^\prime\beta^\prime}\delta_{\beta\gamma^\prime}\delta_{\delta\delta^\prime}(U_{\alpha}U_{\gamma}U_{\alpha^\prime}U_{\beta^\prime}+V_{\alpha}V_{\gamma}V_{\alpha^\prime}V_{\beta^\prime})\notag\\
		-&V_{\alpha\gamma\gamma^\prime\delta^\prime}\delta_{\beta\alpha^\prime}\delta_{\delta\beta^\prime}(U_{\alpha}U_{\gamma}U_{\gamma^\prime}U_{\delta^\prime}+V_{\alpha}V_{\gamma}V_{\gamma^\prime}V_{\delta^\prime})\notag\\
		+&V_{\alpha\gamma\alpha^\prime\gamma^\prime}\delta_{\beta\beta^\prime}\delta_{\delta\delta^\prime}(U_{\alpha}U_{\gamma}U_{\alpha^\prime}U_{\gamma^\prime}+V_{\alpha}V_{\gamma}V_{\alpha^\prime}V_{\gamma^\prime})\notag\\
		+&V_{\alpha\gamma\beta^\prime\delta^\prime}\delta_{\beta\alpha^\prime}\delta_{\delta\gamma^\prime}(U_{\alpha}U_{\gamma}U_{\beta^\prime}U_{\delta^\prime}+V_{\alpha}V_{\gamma}V_{\beta^\prime}V_{\delta^\prime})\notag\\
		-&V_{\alpha\gamma\alpha^\prime\delta^\prime}\delta_{\beta\beta^\prime}\delta_{\delta\gamma^\prime}(U_{\alpha}U_{\gamma}U_{\alpha^\prime}U_{\delta^\prime}+V_{\alpha}V_{\gamma}V_{\alpha^\prime}V_{\delta^\prime})\notag\\
		-&V_{\alpha\gamma\beta^\prime\gamma^\prime}\delta_{\beta\alpha^\prime}\delta_{\delta\delta^\prime}(U_{\alpha}U_{\gamma}U_{\beta^\prime}U_{\gamma^\prime}+V_{\alpha}V_{\gamma}V_{\beta^\prime}V_{\gamma^\prime}) \quad \}\notag\\
		+&(1-P_{\delta\alpha})\{ \notag\\
		&V_{\beta\gamma\alpha^\prime\beta^\prime}\delta_{\alpha\gamma^\prime}\delta_{\delta\delta^\prime}(U_{\beta}U_{\gamma}U_{\alpha^\prime}U_{\beta^\prime}+V_{\beta}V_{\gamma}V_{\alpha^\prime}V_{\beta^\prime})\notag\\
		+&V_{\beta\gamma\gamma^\prime\delta^\prime}\delta_{\alpha\alpha^\prime}\delta_{\delta\beta^\prime}(U_{\beta}U_{\gamma}U_{\gamma^\prime}U_{\delta^\prime}+V_{\beta}V_{\gamma}V_{\gamma^\prime}V_{\delta^\prime})\notag\\
		-&V_{\beta\gamma\alpha^\prime\gamma^\prime}\delta_{\alpha\beta^\prime}\delta_{\delta\delta^\prime}(U_{\beta}U_{\gamma}U_{\alpha^\prime}U_{\gamma^\prime}+V_{\beta}V_{\gamma}V_{\alpha^\prime}V_{\gamma^\prime})\notag\\
		-&V_{\beta\gamma\beta^\prime\delta^\prime}\delta_{\alpha\alpha^\prime}\delta_{\delta\gamma^\prime}(U_{\beta}U_{\gamma}U_{\beta^\prime}U_{\delta^\prime}+V_{\beta}V_{\gamma}V_{\beta^\prime}V_{\delta^\prime})\notag\\
		+&V_{\beta\gamma\alpha^\prime\delta^\prime}\delta_{\alpha\beta^\prime}\delta_{\delta\gamma^\prime}(U_{\beta}U_{\gamma}U_{\alpha^\prime}U_{\delta^\prime}+V_{\beta}V_{\gamma}V_{\alpha^\prime}V_{\delta^\prime})\notag\\
		+&V_{\beta\gamma\beta^\prime\gamma^\prime}\delta_{\alpha\alpha^\prime}\delta_{\delta\delta^\prime}(U_{\beta}U_{\gamma}U_{\beta^\prime}U_{\gamma^\prime}+V_{\beta}V_{\gamma}V_{\beta^\prime}V_{\gamma^\prime})  \quad \}\notag\\
		+&(1-P_{\gamma\beta})\{ \notag\\
		&V_{\alpha\delta\alpha^\prime\beta^\prime}\delta_{\beta\gamma^\prime}\delta_{\gamma\delta^\prime}(U_{\alpha}U_{\delta}U_{\alpha^\prime}U_{\beta^\prime}+V_{\alpha}V_{\delta}V_{\alpha^\prime}V_{\beta^\prime})\notag\\
		+&V_{\alpha\delta\gamma^\prime\delta^\prime}\delta_{\beta\alpha^\prime}\delta_{\gamma\beta^\prime}(U_{\alpha}U_{\delta}U_{\gamma^\prime}U_{\delta^\prime}+V_{\alpha}V_{\delta}V_{\gamma^\prime}V_{\delta^\prime})\notag\\
		-&V_{\alpha\delta\alpha^\prime\gamma^\prime}\delta_{\beta\beta^\prime}\delta_{\gamma\delta^\prime}(U_{\alpha}U_{\delta}U_{\alpha^\prime}U_{\gamma^\prime}+V_{\alpha}V_{\delta}V_{\alpha^\prime}V_{\gamma^\prime})\notag\\
		-&V_{\alpha\delta\beta^\prime\delta^\prime}\delta_{\beta\alpha^\prime}\delta_{\gamma\gamma^\prime}(U_{\alpha}U_{\delta}U_{\beta^\prime}U_{\delta^\prime}+V_{\alpha}V_{\delta}V_{\beta^\prime}V_{\delta^\prime})\notag\\
		+&V_{\alpha\delta\alpha^\prime\delta^\prime}\delta_{\beta\beta^\prime}\delta_{\gamma\gamma^\prime}(U_{\alpha}U_{\delta}U_{\alpha^\prime}U_{\delta^\prime}+V_{\alpha}V_{\delta}V_{\alpha^\prime}V_{\delta^\prime})\notag\\
		+&V_{\alpha\delta\beta^\prime\gamma^\prime}\delta_{\beta\alpha^\prime}\delta_{\gamma\delta^\prime}(U_{\alpha}U_{\delta}U_{\beta^\prime}U_{\gamma^\prime}+V_{\alpha}V_{\delta}V_{\beta^\prime}V_{\gamma^\prime})\quad \}
\end{align}

\begin{align}\label{eqA22v2}
&A_{\alpha\beta\gamma\delta,\alpha^\prime\beta^\prime\gamma^\prime\delta^\prime}^{V2}\notag\\
		=&(1-P_{\delta\gamma})\{\notag\\
		&\delta_{\gamma\gamma^\prime}\delta_{\delta\delta^\prime}(1-P_{\alpha\beta}-P_{\alpha^\prime\beta^\prime}+P_{\alpha\beta}P_{\alpha^\prime\beta^\prime})V_{\alpha\bar{\beta^\prime}\bar{\beta}\alpha^\prime}U_{\alpha}V_{\beta^\prime}V_{\beta}U_{\alpha^\prime}\notag\\
		+&\delta_{\gamma\alpha^\prime}\delta_{\delta\beta^\prime}(1-P_{\alpha\beta}-P_{\gamma^\prime\delta^\prime}+P_{\alpha\beta}P_{\gamma^\prime\delta^\prime})V_{\alpha\bar{\delta^\prime}\bar{\beta}\gamma^\prime}U_{\alpha}V_{\delta^\prime}V_{\beta}U_{\gamma^\prime} \notag\\
		-&\delta_{\gamma\beta^\prime}\delta_{\delta\delta^\prime}(1-P_{\alpha\beta}-P_{\alpha^\prime\gamma^\prime}+P_{\alpha\beta}P_{\alpha^\prime\gamma^\prime})V_{\alpha\bar{\gamma^\prime}\bar{\beta}\alpha^\prime}U_{\alpha}V_{\gamma^\prime}V_{\beta}U_{\alpha^\prime}\notag\\
		-&\delta_{\gamma\alpha^\prime}\delta_{\delta\gamma^\prime}(1-P_{\alpha\beta}-P_{\beta^\prime\delta^\prime}+P_{\alpha\beta}P_{\beta^\prime\delta^\prime})V_{\alpha\bar{\delta^\prime}\bar{\beta}\beta^\prime}U_{\alpha}V_{\delta^\prime}V_{\beta}U_{\beta^\prime}\notag\\
		+&\delta_{\gamma\beta^\prime}\delta_{\delta\gamma^\prime}(1-P_{\alpha\beta}-P_{\alpha^\prime\delta^\prime}+P_{\alpha\beta}P_{\alpha^\prime\delta^\prime})V_{\alpha\bar{\delta^\prime}\bar{\beta}\alpha^\prime}U_{\alpha}V_{\delta^\prime}V_{\beta}U_{\alpha^\prime}\notag\\
		+&\delta_{\gamma\alpha^\prime}\delta_{\delta\delta^\prime}(1-P_{\alpha\beta}-P_{\beta^\prime\gamma^\prime}+P_{\alpha\beta}P_{\beta^\prime\gamma^\prime})V_{\alpha\bar{\gamma^\prime}\bar{\beta}\beta^\prime}U_{\alpha}V_{\gamma^\prime}V_{\beta}U_{\beta^\prime} \}\notag\\
		+&(1-P_{\alpha\beta})\{ \notag\\ &\delta_{\alpha\gamma^\prime}\delta_{\beta\delta^\prime}(1-P_{\gamma\delta}-P_{\alpha^\prime\beta^\prime}+P_{\gamma\delta}P_{\alpha^\prime\beta^\prime})V_{\gamma\bar{\beta^\prime}\bar{\delta}\alpha^\prime}U_{\gamma}V_{\beta^\prime}V_{\delta}U_{\alpha^\prime}\notag\\
		+&\delta_{\alpha\alpha^\prime}\delta_{\beta\beta^\prime}(1-P_{\gamma\delta}-P_{\gamma^\prime\delta^\prime}+P_{\gamma\delta}P_{\gamma^\prime\delta^\prime})V_{\gamma\bar{\delta^\prime}\bar{\delta}\gamma^\prime}U_{\gamma}V_{\delta^\prime}V_{\delta}U_{\gamma^\prime}\notag\\
		-&\delta_{\alpha\beta^\prime}\delta_{\beta\delta^\prime}(1-P_{\gamma\delta}-P_{\alpha^\prime\gamma^\prime}+P_{\gamma\delta}P_{\alpha^\prime\gamma^\prime})V_{\gamma\bar{\gamma^\prime}\bar{\delta}\alpha^\prime}U_{\gamma}V_{\gamma^\prime}V_{\delta}U_{\alpha^\prime}\notag\\
		-&\delta_{\alpha\alpha^\prime}\delta_{\beta\gamma^\prime}(1-P_{\gamma\delta}-P_{\beta^\prime\delta^\prime}+P_{\gamma\delta}P_{\beta^\prime\delta^\prime})V_{\gamma\bar{\delta^\prime}\bar{\delta}\beta^\prime}U_{\gamma}V_{\delta^\prime}V_{\delta}U_{\beta^\prime}\notag\\
		+&\delta_{\alpha\beta^\prime}\delta_{\beta\gamma^\prime}(1-P_{\gamma\delta}-P_{\alpha^\prime\delta^\prime}+P_{\gamma\delta}P_{\alpha^\prime\delta^\prime})V_{\gamma\bar{\delta^\prime}\bar{\delta}\alpha^\prime}U_{\gamma}V_{\delta^\prime}V_{\delta}U_{\alpha^\prime}\notag\\
		+&\delta_{\alpha\alpha^\prime}\delta_{\beta\delta^\prime}(1-P_{\gamma\delta}-P_{\beta^\prime\gamma^\prime}+P_{\gamma\delta}P_{\beta^\prime\gamma^\prime})V_{\gamma\bar{\gamma^\prime}\bar{\delta}\beta^\prime}U_{\gamma}V_{\gamma^\prime}V_{\delta}U_{\beta^\prime} \}\notag\\
		+&(1-P_{\alpha\gamma})\{\notag\\
		-&\delta_{\alpha\gamma^\prime}\delta_{\gamma\delta^\prime}(1-P_{\beta\delta}-P_{\alpha^\prime\beta^\prime}+P_{\beta\delta}P_{\alpha^\prime\beta^\prime})V_{\beta\bar{\beta^\prime}\bar{\delta}\alpha^\prime}U_{\beta}V_{\beta^\prime}V_{\delta}U_{\alpha^\prime}\notag\\
		-&\delta_{\alpha\alpha^\prime}\delta_{\gamma\beta^\prime}(1-P_{\beta\delta}-P_{\gamma^\prime\delta^\prime}+P_{\beta\delta}P_{\gamma^\prime\delta^\prime})V_{\beta\bar{\delta^\prime}\bar{\delta}\gamma^\prime}U_{\beta}V_{\delta^\prime}V_{\delta}U_{\gamma^\prime}\notag\\
		+&\delta_{\alpha\beta^\prime}\delta_{\gamma\delta^\prime}(1-P_{\beta\delta}-P_{\alpha^\prime\gamma^\prime}+P_{\beta\delta}P_{\alpha^\prime\gamma^\prime})V_{\beta\bar{\gamma^\prime}\bar{\delta}\alpha^\prime}U_{\beta}V_{\gamma^\prime}V_{\delta}U_{\alpha^\prime}\notag\\
		+&\delta_{\alpha\alpha^\prime}\delta_{\gamma\gamma^\prime}(1-P_{\beta\delta}-P_{\beta^\prime\delta^\prime}+P_{\beta\delta}P_{\beta^\prime\delta^\prime})V_{\beta\bar{\delta^\prime}\bar{\delta}\beta^\prime}U_{\beta}V_{\delta^\prime}V_{\delta}U_{\beta^\prime}\notag\\
		-&\delta_{\alpha\beta^\prime}\delta_{\gamma\gamma^\prime}(1-P_{\beta\delta}-P_{\alpha^\prime\delta^\prime}+P_{\beta\delta}P_{\alpha^\prime\delta^\prime})V_{\beta\bar{\delta^\prime}\bar{\delta}\alpha^\prime}U_{\beta}V_{\delta^\prime}V_{\delta}U_{\alpha^\prime}\notag\\
		-&\delta_{\alpha\alpha^\prime}\delta_{\gamma\delta^\prime}(1-P_{\beta\delta}-P_{\beta^\prime\gamma^\prime}+P_{\beta\delta}P_{\beta^\prime\gamma^\prime})V_{\beta\bar{\gamma^\prime}\bar{\delta}\beta^\prime}U_{\beta}V_{\gamma^\prime}V_{\delta}U_{\beta^\prime} \}\notag\\
		+&(1-P_{\delta\beta})\{\notag\\
		-&\delta_{\beta\gamma^\prime}\delta_{\delta\delta^\prime}(1-P_{\alpha\gamma}-P_{\alpha^\prime\beta^\prime}+P_{\alpha\gamma}P_{\alpha^\prime\beta^\prime})V_{\alpha\bar{\beta^\prime}\bar{\gamma}\alpha^\prime}U_{\alpha}V_{\beta^\prime}V_{\gamma}U_{\alpha^\prime}\notag\\
		-&\delta_{\beta\alpha^\prime}\delta_{\delta\beta^\prime}(1-P_{\alpha\gamma}-P_{\gamma^\prime\delta^\prime}+P_{\alpha\gamma}P_{\gamma^\prime\delta^\prime})V_{\alpha\bar{\delta^\prime}\bar{\gamma}\gamma^\prime}U_{\alpha}V_{\delta^\prime}V_{\gamma}U_{\gamma^\prime}\notag\\
		+&\delta_{\beta\beta^\prime}\delta_{\delta\delta^\prime}(1-P_{\alpha\gamma}-P_{\alpha^\prime\gamma^\prime}+P_{\alpha\gamma}P_{\alpha^\prime\gamma^\prime})V_{\alpha\bar{\gamma^\prime}\bar{\gamma}\alpha^\prime}U_{\alpha}V_{\gamma^\prime}V_{\gamma}U_{\alpha^\prime}\notag\\
		+&\delta_{\beta\alpha^\prime}\delta_{\delta\gamma^\prime}(1-P_{\alpha\gamma}-P_{\beta^\prime\delta^\prime}+P_{\alpha\gamma}P_{\beta^\prime\delta^\prime})V_{\alpha\bar{\delta^\prime}\bar{\gamma}\beta^\prime}U_{\alpha}V_{\delta^\prime}V_{\gamma}U_{\beta^\prime}\notag\\
		-&\delta_{\beta\beta^\prime}\delta_{\delta\gamma^\prime}(1-P_{\alpha\gamma}-P_{\alpha^\prime\delta^\prime}+P_{\alpha\gamma}P_{\alpha^\prime\delta^\prime})V_{\alpha\bar{\delta^\prime}\bar{\gamma}\alpha^\prime}U_{\alpha}V_{\delta^\prime}V_{\gamma}U_{\alpha^\prime}\notag\\
		-&\delta_{\beta\alpha^\prime}\delta_{\delta\delta^\prime}(1-P_{\alpha\gamma}-P_{\beta^\prime\gamma^\prime}+P_{\alpha\gamma}P_{\beta^\prime\gamma^\prime})V_{\alpha\bar{\gamma^\prime}\bar{\gamma}\beta^\prime}U_{\alpha}V_{\gamma^\prime}V_{\gamma}U_{\beta^\prime} \}\notag\\
		+&(1-P_{\delta\alpha})\{ \notag\\ &\delta_{\alpha\gamma^\prime}\delta_{\delta\delta^\prime}(1-P_{\beta\gamma}-P_{\alpha^\prime\beta^\prime}+P_{\beta\gamma}P_{\alpha^\prime\beta^\prime})V_{\beta\bar{\beta^\prime}\bar{\gamma}\alpha^\prime}U_{\beta}V_{\beta^\prime}V_{\gamma}U_{\alpha^\prime}\notag\\
		+&\delta_{\alpha\alpha^\prime}\delta_{\delta\beta^\prime}(1-P_{\beta\gamma}-P_{\gamma^\prime\delta^\prime}+P_{\beta\gamma}P_{\gamma^\prime\delta^\prime})V_{\beta\bar{\delta^\prime}\bar{\gamma}\gamma^\prime}U_{\beta}V_{\delta^\prime}V_{\gamma}U_{\gamma^\prime}\notag\\
		-&\delta_{\alpha\beta^\prime}\delta_{\delta\delta^\prime}(1-P_{\beta\gamma}-P_{\alpha^\prime\gamma^\prime}+P_{\beta\gamma}P_{\alpha^\prime\gamma^\prime})V_{\beta\bar{\gamma^\prime}\bar{\gamma}\alpha^\prime}U_{\beta}V_{\gamma^\prime}V_{\gamma}U_{\alpha^\prime}\notag\\
		-&\delta_{\alpha\alpha^\prime}\delta_{\delta\gamma^\prime}(1-P_{\beta\gamma}-P_{\beta^\prime\delta^\prime}+P_{\beta\gamma}P_{\beta^\prime\delta^\prime})V_{\beta\bar{\delta^\prime}\bar{\gamma}\beta^\prime}U_{\beta}V_{\delta^\prime}V_{\gamma}U_{\beta^\prime}\notag\\
		+&\delta_{\alpha\beta^\prime}\delta_{\delta\gamma^\prime}(1-P_{\beta\gamma}-P_{\alpha^\prime\delta^\prime}+P_{\beta\gamma}P_{\alpha^\prime\delta^\prime})V_{\beta\bar{\delta^\prime}\bar{\gamma}\alpha^\prime}U_{\beta}V_{\delta^\prime}V_{\gamma}U_{\alpha^\prime}\notag\\
		+&\delta_{\alpha\alpha^\prime}\delta_{\delta\delta^\prime}(1-P_{\beta\gamma}-P_{\beta^\prime\gamma^\prime}+P_{\beta\gamma}P_{\beta^\prime\gamma^\prime})V_{\beta\bar{\gamma^\prime}\bar{\gamma}\beta^\prime}U_{\beta}V_{\gamma^\prime}V_{\gamma}U_{\beta^\prime} \}\notag\\
		+&(1-P_{\gamma\beta})\{\notag\\
		&\delta_{\beta\gamma^\prime}\delta_{\gamma\delta^\prime}(1-P_{\alpha\delta}-P_{\alpha^\prime\beta^\prime}+P_{\alpha\delta}P_{\alpha^\prime\beta^\prime})V_{\alpha\bar{\beta^\prime}\bar{\delta}\alpha^\prime}U_{\alpha}V_{\beta^\prime}V_{\delta}U_{\alpha^\prime}\notag\\
		+&\delta_{\beta\alpha^\prime}\delta_{\gamma\beta^\prime}(1-P_{\alpha\delta}-P_{\gamma^\prime\delta^\prime}+P_{\alpha\delta}P_{\gamma^\prime\delta^\prime})V_{\alpha\bar{\delta^\prime}\bar{\delta}\gamma^\prime}U_{\alpha}V_{\delta^\prime}V_{\delta}U_{\gamma^\prime}\notag\\
		-&\delta_{\beta\beta^\prime}\delta_{\gamma\delta^\prime}(1-P_{\alpha\delta}-P_{\alpha^\prime\gamma^\prime}+P_{\alpha\delta}P_{\alpha^\prime\gamma^\prime})V_{\alpha\bar{\gamma^\prime}\bar{\delta}\alpha^\prime}U_{\alpha}V_{\gamma^\prime}V_{\delta}U_{\alpha^\prime}\notag\\
		-&\delta_{\beta\alpha^\prime}\delta_{\gamma\gamma^\prime}(1-P_{\alpha\delta}-P_{\beta^\prime\delta^\prime}+P_{\alpha\delta}P_{\beta^\prime\delta^\prime})V_{\alpha\bar{\delta^\prime}\bar{\delta}\beta^\prime}U_{\alpha}V_{\delta^\prime}V_{\delta}U_{\beta^\prime}\notag\\
		+&\delta_{\beta\beta^\prime}\delta_{\gamma\gamma^\prime}(1-P_{\alpha\delta}-P_{\alpha^\prime\delta^\prime}+P_{\alpha\delta}P_{\alpha^\prime\delta^\prime})V_{\alpha\bar{\delta^\prime}\bar{\delta}\alpha^\prime}U_{\alpha}V_{\delta^\prime}V_{\delta}U_{\alpha^\prime}\notag\\
		+&\delta_{\beta\alpha^\prime}\delta_{\gamma\delta^\prime}(1-P_{\alpha\delta}-P_{\beta^\prime\gamma^\prime}+P_{\alpha\delta}P_{\beta^\prime\gamma^\prime})V_{\alpha\bar{\gamma^\prime}\bar{\delta}\beta^\prime}U_{\alpha}V_{\gamma^\prime}V_{\delta}U_{\beta^\prime} \}
\end{align}

\begin{equation}\label{eqB12B22}
	\begin{split}
		&B_{24}= B_{\alpha\beta,\alpha^\prime\beta^\prime\gamma^\prime\delta^\prime}=-<|[\alpha_\beta\alpha_\alpha,H,\alpha_{\delta^\prime}\alpha_{\gamma^\prime}\alpha_{\beta^\prime}\alpha_{\alpha^\prime}]|>=0\\
		&B_{44}=B_{\alpha\beta\gamma\delta,\alpha^\prime\beta^\prime\gamma^\prime\delta^\prime}=-<|[\alpha_\delta\alpha_\gamma\alpha_\beta\alpha_\alpha,H,\alpha_{\delta^\prime}\alpha_{\gamma^\prime}\alpha_{\beta^\prime}\alpha_{\alpha^\prime}]|>=0
	\end{split}
\end{equation}

The quantities $E_\alpha$, $E_\beta$, $E_\gamma$, and $E_\delta$ in the energy terms of $A_{22}$ and $A_{44}$ denote the corresponding quasiparticle energies. The $\rm V^{PP}$ in $\rm A_{22}$ and $\rm B_{22}$ denotes the particle-particle residual-interaction channel. Here, $P_{ij}$ denotes the operator that exchanges subscripts $i$ and $j$. 
The matrix elements $B_{24}$ and $B_{44}$ vanish because the double commutators evaluated with respect to the quasiparticle vacuum yield zero by construction.

\section{{II.  Necessity of the subtraction procedure}}
\begin{figure}[!ht]
\centering
\includegraphics[width=\linewidth]{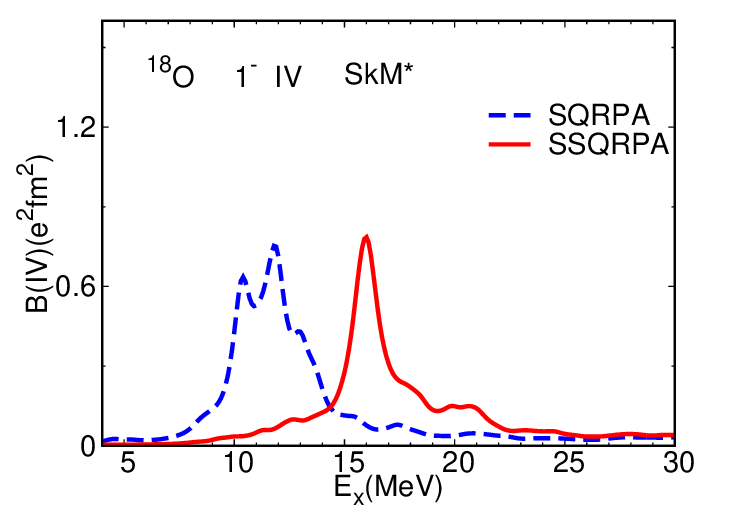}
\caption{IV E1 strength distributions in $^{18}$O calculated with SQRPA excluding (blue dashed line) and including (red line) subtraction procedure, respectively 
.}
\label{fig:sub}
\end{figure}

The subtraction procedure modifies the SQRPA matrices above by removing spurious contributions from ground-state correlations already encoded in the energy-density functional~\cite{PhysRevC.88.054301,PhysRevC.92.034303}. Its necessity is illustrated in Fig.~\ref{fig:sub}.
The subtraction procedure plays a crucial role in the  SQRPA calculations; therefore, we provide an illustrative  example and discussion here.
As shown in Fig.~\ref{fig:sub}, the strength distributions are  significantly shifted downwards with the SQRPA  calculation without the  subtraction procedure, compared to the result obtained with the  subtraction procedure. Furthermore, the calculation without the subtraction procedure can yield complex eigenfrequencies, indicating unstable solutions. Consequently, SQRPA calculations without the subtraction procedure are not considered in this work.

\bibliography{E1}